\documentclass[a4paper,fleqn]{cas-sc}

\usepackage[authoryear,longnamesfirst]{natbib}

\def\tsc#1{\csdef{#1}{\textsc{\lowercase{#1}}\xspace}}
\tsc{WGM}
\tsc{QE}
\tsc{EP}
\tsc{PMS}
\tsc{BEC}
\tsc{DE}
\NewDocumentEnvironment { Abstract } { o }{}

\begin{document}
\let\WriteBookmarks\relax
\def\floatpagepagefraction{1}
\def\textpagefraction{.001}

\shorttitle{From Consciousness to Society}

\shortauthors{M. Albarracin}

\title [mode = title]{Comment on ``Path integrals, Particular Kinds, and Strange Things'': from Consciousness to Society}                      
% \tnotemark[1,2]

% \tnotetext[1]{This document is the results of the research
%   project funded by the National Science Foundation.}

% \tnotetext[2]{The second title footnote which is a longer text matter
%   to fill through the whole text width and overflow into
%   another line in the footnotes area of the first page.}

\author[1,2]{Mahault Albarracin}

\cormark[1]

% % Footnote of the first author

\ead{albarracin.mahault@courrier.uqam.ca}

% \ead[url]{www.cvr.cc, cvr@sayahna.org}

% \credit{Conceptualization of this study, Methodology, Software}

\affiliation[1]{organization={Université du Québec a Montréal, Faculté des arts et des sciences - Département d’informatique et de recherche opérationnelle, },
    addressline={3150 Jean Brillant St}, 
    city={Montreal},
    % citysep={}, % Uncomment if no comma needed between city and postcode
    postcode={H3T 1N8}, 
    country={Quebec, Canada}}

% \author[2,4]{Riddhi J. Pitliya}[style=chinese]

% % Third author
% \author[2,3]{CV Rajagopal}[%
%   role=Co-ordinator,
%   suffix=Jr,
% \ead{cvr3@sayahna.org}
% \ead[URL]{www.sayahna.org}

% \credit{Data curation, Writing - Original draft preparation}

\affiliation[2]{organization={VERSES},
    city={Los Angeles},
    % citysep={}, % Uncomment if no comma needed between city and postcode
    % postcode={695014}, 
    state={California},
    country={USA}}

% \author%
% [1,3]
% {Rishi T.}
% \cormark[2]
% \fnmark[1,3]
% \ead{rishi@stmdocs.in}
% \ead[URL]{www.stmdocs.in}

% \affiliation[3]{organization={Department of Experimental Psychology, University of Oxford},
%     addressline={Oxford}, 
%     % city={Malayinkil},
%     % citysep={}, % Uncomment if no comma needed between city and postcode
%     % postcode={695571}, 
%     % state={Trivandrum},
%     country={ UK}}

\cortext[cor1]{Corresponding author}
% \cortext[cor2]{Principal corresponding author}

% % Footnote text
% \fntext[fn1]{This is the first author footnote. but is common to third
% \fntext[fn2]{Another author footnote, this is a very long footnote and
%   it should be a really long footnote. But this footnote is not yet
%   sufficiently long enough to make two lines of footnote text.}

% \nonumnote{This note has no numbers. In this work we demonstrate $a_b$
%   the formation Y\_1 of a new type of polariton on the interface
%   between a cuprous oxide slab and a polystyrene micro-sphere placed

% \begin{abstract}
% The authors present some challenges to the core claims of the FEP. They refer to the strategy of the “Markov blanket trick”. In this commentary, we respond to some of these challenges. 

% % \noindent\texttt{\textbackslash begin{abstract}} \dots 
% % \texttt{\textbackslash end{abstract}} and
% % \verb+\begin{keyword}+ \verb+...+ \verb+\end{keyword}+ 
% % which
% % contain the abstract and keywords respectively. 

% % \noindent Each keyword shall be separated by a \verb+\sep+ command.
% \end{abstract}

% \includegraphics{figs/grabs.pdf}

% % Research highlights
% \begin{highlights}
% \item Research highlights item 1
% \item Research highlights item 2
% \item Research highlights item 3
% \end{highlights}

% Keywords
% \begin{keywords}
% Active Inference \sep Markov Blanket \sep Free energy principle \sep Modeling
% \end{keywords}

\maketitle

\section{Introduction}
I am pleased to comment on ``Path integrals, particular kinds, and strange things'' by Friston and colleagues (2023). The path integral formulation introduced in this paper offers an innovative lens for examining complex systems, particularly through its nuanced typology of particles. This comprehensive commentary focuses on ‘strange particles,’ a distinct category whose concealed active states provide an intriguing model for understanding phenomena like consciousness and resilience \citep{friston2023path}. 
A key contribution of the formulation is its ability to express system dynamics in terms of densities over paths or trajectories, thereby circumventing the need for assumptions about stationary states. This is especially advantageous for analyzing non-stationary systems. The formulation introduces a nuanced typology of particles—active, inert, dissipative, conservative, ordinary, and strange—each with unique properties that offer valuable insights into resilience, or specifically the notion of timescales and possibility for inner screens.

\section{Types of Particles and Resilience}
The paper categorizes particles into various types, each with unique properties \citep{friston2023path}. These particle types map onto the concepts of inertia, elasticity, and plasticity.
Inert particles show pure inertia, simply resisting perturbations using high precision states. Without active states, they cannot change or adapt, thereby embodying a form of resilience that is limited to resistance \citep{miller2022resilience}.
Active particles have internal states influencing external states through active states which allows for elastic relaxation to characteristic states after perturbations, implementing a basic form of homeostasis \citep{miller2022resilience}.
Dissipative particles experience randomness that prevents pure inertia. Their fluctuations create possibilities for plasticity, as entropy is dissipated away from equilibrium \citep{miller2022resilience}.
Conservative particles follow precise paths of least action. Their dynamics display robust inertia alongside active elasticity. Temporally deep models enable planning routes back to preferences.
Ordinary particles learn from the consequences of actions on internal states, exhibiting plasticity by changing beliefs and policies. This functional redundancy affords adaptive possibilities.
Strange particles combine inertia with plasticity by hiding active states. Apparent autonomy arises from inferring actions as hidden causes. Through this process, Strange particles can expand behavioral repertoires, enabling resilience, and ultimately self-organization.

\section{Strange particles, Autonomy and quantifying consciousness}
Strange particles’ concealed active states, which are not directly influenced by internal states but are mediated through external states, grant them an illusion of autonomy or agency. Their internal states seem to infer external and active paths as the hidden causes behind sensory inputs. This characteristic enables us to investigate how phenomena like agency and consciousness might emerge. The nested Markov blanket architecture in strange particles supports information processing and inference, generating seemingly autonomous behavior \citep{ramstead2023inner}.
The concealed active states and nested architecture establish strange particles as candidates for differentiating systems that exhibit signs of conscious processing \cite{friston2023path}. This aligns with the perspective that consciousness is extended to the social realm through linguistic exchange, creating a shared reality. The nested structure can be likened to “inner screens” in more complex systems, with generative models unfolding over slower timescales to contextualize and constrain lower-level inference. Deeply nested strange particles essentially constitute such inner screen models, qualifying as candidates for consciousness.

\section{Embedded Normativity and Timescales}
The Lagrangian measures provided by the target paper, combined with the typology of particles in terms of resilience, and as a measure of conscious processes, give us a framework for quantifying system dynamics across different timescales. While not direct indicators of consciousness, these measures offer discrimination between particles based on properties like active states \citep{friston2023path}. The Lagrangian also scores the diversity of paths, placing entropy bounds within given timescales. This is particularly relevant for strange particles as candidates for measuring consciousness due to their hidden active states \citep{ramstead2023inner}.

From the perspective of an observer operating on very fast timescales, inert particles may appear static but demonstrate microscopic fluctuations at appropriately sensitive timescales. An observer capable of detecting subtle biochemical fluctuations could identify inert dynamics at microscopic scales. Conversely, an observer tuned to circadian or seasonal periodicities could discern the habitual nature of elastic particles, such as active and dissipative particles, which support elastic resilience through homeostasis and allostasis. These particles enable the system to return to characteristic states after perturbations \citep{miller2022resilience}.

Ultimately, strange particles facilitate larger-scale self-organization across various timescales, integrating with other particles and dynamics as part of a multiscale architecture. Lower inertial dynamics supply variability that higher adaptive levels integrate through niche construction \citep{guenin2023embedded}. This highlights the nested multiscale constraints linking biochemistry to physiology to behavior in the emergence of autonomy.

Embedded normativity provides a natural operational concept for the study of social and ecological robotics across scales of activity. It emerges from the multiscale integration of social and cognitive constraints occurring through engagement with the world. The constraints of the agent’s environment are as meaningful carriers of normativity as the constraints of the agent’s organism. This paints a radically counter-intuitive picture where normativity exists not inside or outside agents, but through statistical engagement properties.

At extended timescales spanning generations, the repertoire expansion of ordinary and strange particles manifests as plasticity, shaping slowly evolving social norms. These particles exhibit a behavioral flexibility and capacity for novelty that allows them to optimize precision and learn over time. Detecting this plasticity would require an observer with a long-term perspective across multiple lifetimes.

The timescales and properties of particles are relative to the observer's perspective. Across multiple spatiotemporal frames, the continuum of inertial, elastic, and plastic dynamics scaffolds the embedded normativity that guides self-organization. A complete understanding requires integrating observations across scales.

\section{Conclusion}
The path integral perspective offers a compelling physics of life that could formally quantify and qualify conscious and larger scale processes. As we map inertia, elasticity, and plasticity onto the states, dependencies, and scales of particles, the conjunction of these fields opens a promising direction to understand the embodied emergence of adaptive autonomy. Through these formulations, we hope we can find a path to furthering our goal of a unified framework for studying complex, adaptive systems.

\section{Acknowledgement}
I wish to thank Maxwell Ramstead and Lance Da Costa for the discussions that led to an improved version of this comment. 

% \appendix
% \section{My Appendix}
% Appendix sections are coded under \verb+\appendix+.

% \verb+\printcredits+ command is used after appendix sections to list 
% author credit taxonomy contribution roles tagged using \verb+\credit+ 
% in frontmatter.


\bibliographystyle{cas-model2-names}

\bibliography{cas-refs}


% Author biography without author photo.
% Author biography. Author biography. Author biography.
% Author biography. Author biography. Author biography.
% Author biography. Author biography. Author biography.
% Author biography. Author biography. Author biography.
% Author biography. Author biography. Author biography.
% Author biography. Author biography. Author biography.
% Author biography. Author biography. Author biography.
% Author biography. Author biography. Author biography.
% Author biography. Author biography. Author biography.

% \bio{figs/pic1}
% Author biography with author photo.
% Author biography. Author biography. Author biography.
% Author biography. Author biography. Author biography.
% Author biography. Author biography. Author biography.
% Author biography. Author biography. Author biography.
% Author biography. Author biography. Author biography.
% Author biography. Author biography. Author biography.
% Author biography. Author biography. Author biography.
% Author biography. Author biography. Author biography.
% Author biography. Author biography. Author biography.

% \bio{figs/pic1}
% Author biography with author photo.
% Author biography. Author biography. Author biography.
% Author biography. Author biography. Author biography.
% Author biography. Author biography. Author biography.
% Author biography. Author biography. Author biography.

\end{document}